\documentclass[journal=jacsat,manuscript=article,layout=twocolumn]{achemso}
\usepackage[version=3]{mhchem} % Formula subscripts using \ce{}
\usepackage[utf8]{inputenc}
\usepackage[T1]{fontenc}   % enables \k (ogonek) etc. in text fonts
\usepackage{lmodern}       % modern T1 fonts (recommended)
\usepackage{textcomp}      % \textminus
\DeclareUnicodeCharacter{2212}{\textminus}  % map U+2212 to a text minus
\usepackage{changes}
\usepackage{gensymb} % For \degree
\usepackage{xr} % For cross-referencing
\usepackage{subcaption}
\makeatletter
\newcommand*{\addFileDependency}[1]{% argument=file name and extension
\typeout{(#1)}% latexmk will find this if $recorder=0
\@addtofilelist{#1}
\IfFileExists{#1}{}{\typeout{No file #1.}}
}\makeatother

\usepackage{url}
\usepackage[hidelinks]{hyperref}% add hypertext capabilities
\author{Rashid Rafeek V Valappil}
\affiliation[IISER]{Department of Chemistry, Indian Institute of Science Education and Research Bhopal, Bhopal 462 066, India}

\author{Sayan Maity}
\affiliation[IISER]{Department of Chemistry, Indian Institute of Science Education and Research Bhopal, Bhopal 462 066, India}
\altaffiliation{Present address: Center for Theoretical Chemistry, Ruhr-Universitaet Bochum, Bochum, Germany.}

\author{Varadharajan Srinivasan}
\email{vardha@iiserb.ac.in}
\affiliation[IISER]{Department of Chemistry, Indian Institute of Science Education and Research Bhopal, Bhopal 462 066, India}

\title[An \textsf{achemso} demo]
  {Elucidating the High-Pressure Phases of MAPbBr\textsubscript{3} Using a Machine Learning Force Field\\}

\abbreviations{MA,GMM,RDF}


\begin{document}

%%%%%%%%%%%%%%%%%%%%%%%%%%%%%%%%%%%%%%%%%%%%%%%%%%%%%%%%%%%%%%

\begin{abstract}
High-pressure phases of the hybrid perovskite \ce{MAPbBr3} have been investigated in detail using a novel machine learning force field (MLFF). MLFF simulations successfully reproduce the sequence of pressure-induced phase transitions from the $\alpha$ ($Pm\bar{3}m$) to the $\beta$ ($Im\bar{3}$) and finally the $\gamma$ ($Pnma$/$Pmn2_1$) phase. In the $\alpha$ phase, the simulations confirm the triple-well character of the potential energy surface for octahedral tilting shedding light into the local dynamic distortions. In the $\beta$ phase, our simulations reveal MA sublattice doubling yielding both orientationally disordered and ordered MA ions mirroring experimental observation. This mixed-order phase results from locally frustrated host-guest couplings arising from the in-phase octahedral tilt system ($a^+a^+a^+$). In the high-pressure $\gamma$ phase, we confirm the formation of polar and anti-polar domains, with the latter have higher lifetimes and persist for over 50 ps at pressures above 1.5 GPa. By elucidating the behavior of various phases of \ce{MAPbBr3}, this work provides a fundamental understanding of how host-guest interactions and octahedral tilting govern the material's properties. Further, the importance of time scales and length scales in characterizing these phases is emphasized.
\end{abstract}

\maketitle

\section{Introduction}
Hybrid perovskites exhibit remarkable photovoltaic properties and have in turn led to the development of materials that show state-of-the-art photovoltaic performance, such as tandem solar cells\cite{dewolfTandems2023}. Pressure can be used to enhance the properties of these materials by bandgap engineering, modifying carrier recombinations, enhancing stability etc. Methylammonium (MA, \ce{CH3NH3}) lead bromide perovskite is one of the materials which has recieved significant attention due to its photovolotaic efficiency. 

Pressure-induced studies on \ce{MAPbBr3} have shown that a transition occurs from the ambient pressure $\alpha$ phase with cubic space group $Pm\bar{3}m$ to another cubic $\beta$ phase with space group $Im\bar{3}$ in the pressure range of 0.4 GPa to 1 GPa\cite{swainsonPressureResponseOrganic2007,wangPressureInducedPhaseTransformation2015a,jaffeHighPressureSingleCrystalStructures2016,capitaniLockingMethylammoniumPressureEnhanced2017,zhangEffectsNonhydrostaticStress2017,yinHighPressureInduced2018,yesudhasCouplingOrganicCation2020,liangReassigningPressureInducedPhase2022}. At higher pressures a second transition occurs to an orthorhombic $\gamma$ phase with space group $Pnma$, in the pressure range 1.5 GPa -- 2.7 GPa\cite{wangPressureInducedPhaseTransformation2015a,zhangEffectsNonhydrostaticStress2017,yinHighPressureInduced2018,yesudhasCouplingOrganicCation2020}. Also, a mixture of the $\beta$ and the $\gamma$ has been reported to form at lower pressures of 1.1 GPa when \ce{Ar} was used as the pressure-transmitting medium (PTM)\cite{zhangEffectsNonhydrostaticStress2017}. The space group assignment of the high pressure $\gamma$ phase, which is non-polar, was disputed by a later report, which assigned the space group to a polar $Pmn2_1$\cite{liangReassigningPressureInducedPhase2022}. Although most studies report amorphization at even higher pressures, a recent report suggests the existence of an additional crystalline phase beyond 4.6 GPa with the $Pmn2_1$ space group and remains stable up to at least 7.6 GPa\cite{liangPressureInducedPhaseTransition2023}.

% Previously, we carried out ab initio molecular dynamics simulations to investigate the phase transformations in \ce{MAPbBr3} under temperature and pressure\cite{maity_deciphering_2022,maityStabilizingPolarDomains2023,maityCooperativeOctahedralTilt2024a}. The characteristics of the pressure-induced phases as well as These revealed the formation of polar domains in the high-pressure phase. 
Dynamics of the MA (guest) and the octahedral cage (host) in each of these phases have been investigated in several works. 
The ambient pressure cubic $\alpha$ phase features an orientationally disordered MA in an average pseudo-cubic octahedral cage whose average tilts are denoted in glazer notation as $a^0a^0a^0$~\cite{glazerClassificationTiltedOctahedra1972,swainsonPhaseTransitionsPerovskite2003,zhangEffectsNonhydrostaticStress2017}. AIMD simulations conducted by our group, have captured this MA disorder, reproducing the experimentally obtained timescales of MA reorientational motion with good accuracy\cite{maity_deciphering_2022,maityStabilizingPolarDomains2023}. It was found that even though MAs sample the full orientational space, there was a preference for edge-diagonal orientations. The octahedral lattice motion was found to be weakly coupled to the MAs in this phase through N-H...Br and C-H...Br hydrogen bonds (H-bonds). Further, although the ensemble-averaged octahedral tilt	is zero, different MA orientations stabilize both tilted and untilted configurations, resulting in a triple-well PES for the octahedral tilts in the $\alpha$ phase~\cite{maityCooperativeOctahedralTilt2024a}.

The transition to the intermediate pressure $\beta$ phase is associated with a unit cell doubling and a change from $a^0a^0a^0$ to $a^+a^+a^+$ octahedral tilt ordering~\cite{jaffeHighPressureSingleCrystalStructures2016,swainsonPressureResponseOrganic2007,zhangEffectsNonhydrostaticStress2017}. A previous single crystal experiment has reported that this phase hosts two types of MA ions that differ in the type of disorder displayed~\cite{jaffeHighPressureSingleCrystalStructures2016}. This behavior was hypothesized to arise from different H-bonding interactions in the two sites. AIMD simulations have captured this $\alpha$ to $\beta$ transition, marked by sharp changes in Pb-Br bond lengths and Pb-Br-Pb bond angles~\cite{maityStabilizingPolarDomains2023}. The associated ordering of octahedral tilts from multimodal in the $\alpha$ phase to unimodal character further showed an $a^0$ to $a^+$ tilt transition. However, such a transition was only captured weakly along other directions due to a tilt-disorder over layers~\cite{maityCooperativeOctahedralTilt2024a}. This behavior was suggested to be an indication of coexistence of the $\alpha$ and $\beta$ phases, although the AIMD timescales were insufficient to confirm this hypothesis.
 
 The high-pressure orthorhombic $\gamma$ phase is associated with an octahedral lattice represented by $a^+b^-b^-$ tilts and orientational ordering of MA ions~\cite{yinHighPressureInduced2018,capitaniLockingMethylammoniumPressureEnhanced2017}. AIMD results showed that the orientational ordering of MAs in this phase, specifically results in a long-range staggered ordering (LRSO) of MAs in one plane, similar to the one observed in the low temperature $Pnma$ phase~\cite{maity_deciphering_2022,maityStabilizingPolarDomains2023}. Further, a static disorder of MA dipoles is present here, which leads to the formation of polar and anti-polar domains~\cite{maityStabilizingPolarDomains2023}. The primary order parameter for the transitions were identified as the octahedral tilt modes which were found to display large coupling to the guest MAs in the $\gamma$ phase~\cite{maityCooperativeOctahedralTilt2024a}. However, the AIMD simulations indicated a tetragonal phase instead of the experimentally obtained orthorhombic phase.
 
 In spite of the understanding described above, some characteristics of these phases are still unclear. The experimental finding of multiple types of MA ions in the $\beta$ phase~\cite{jaffeHighPressureSingleCrystalStructures2016} was not obtained in previous AIMD simulations~\cite{maityStabilizingPolarDomains2023}, disallowing investigations into its origin. Further, a surprising preference for a specific plane, incompatible with the cubic space group, was seen for the MAs in the $\beta$ phase in this work. This indicates potential time scale issues may have limited the obtained insights. 
Additionally, even though the formation of polar domains were depicted in the $\gamma$ phase, a detailed investigation of its dynamics and domain sizes were limited by the practically achievable time and length scales of AIMD simulations. The disagreement of experiments and simulation on the orthorhombic nature of the $\gamma$ phase needs to explained as well. Recent combined experimetal and simulation study has observed the formation of twinned dynamic nanodomains with finite octahedral tilt in the $\alpha$ phase\cite{dubajic_dynamic_2025}.  
This raises the question of whether such nanodomains are related to the triple-well PES of octahedral tilt modes obtained in AIMD simulations~\cite{maityCooperativeOctahedralTilt2024a}. Formation of such nanodomains can be affected by simulation cell sizes as they were found to span at least 6 unit cells, making larger length scales important for such investigations. 

Addressing these requires larger time-scales and length-scales than was considered in the AIMD simulations. However, the large cost of \textit{ab initio} calculations limits their applicability for larger lengths and longer times.  Machine learning force fields (MLFF) have risen as a possible way to efficiently conduct such simulations without sacrificing on accuracy\cite{zhang_deep_2018}. Although there have been several works utilizing MLFFs to study phase transitions in \ce{MAPbBr3} and similar hybrid halide perovskites\cite{dubajic_dynamic_2025,maczkaPhaseTransitionsDielectric2023,tuo_spontaneous_2023,liang_structural_2023,fykouras_disorder_2023,fransson_revealing_2023,bokdam_exploring_2021,jinnouchi_phase_2019}, study of pressure-induced transformations in such systems are absent.
Herein, we train a MLFF for the simulation of pressure-induced phases of \ce{MAPbBr3}. This MLFF, which was trained using data from previous AIMD simulations and additional structures obtained through concurrent learning,  was used to conduct molecular dynamics simulations to characterize each of the high-pressure phases of \ce{MAPbBr3} using large system sizes and long time scales. The MLFF is found to accurately represent the properties each of $\alpha$, $\beta$ and $\gamma$ phases in comparison to both previous simulations and experiments. The simulations reproduce the order of each of these phases and their characteristics such as the guest MA reorientational distributions and host octahedral tilt behavior. We provide important insights on each of these phases, bridging existing gaps in understanding and expanding on them. The unique MA ordering in the $\beta$ phase is explained as a result of its octahedral tilting and its coupling to the MAs. The importance of simulation times and lengths are emphasized in obtaining the $\beta$ phase accurately. We also highlight other underrepresented characteristics of the phases such as the triple-well PES for octahedral tilts in the $\alpha$ phase and dynamics of polar domains in the $\gamma$ phase.

\section{Computational methods} 

\subsection{MLFF Training}

A machine learning force field (MLFF) was trained using the DeepMD formalism using data obtained from Car–Parrinello molecular dynamics\cite{carUnifiedApproachMolecular1985} (CPMD) simulations carried out using the Quantum ESPRESSO code\cite{giannozziQUANTUMESPRESSOModular2009}. More details on the method used to generate the \textit{ab initio} data are provided in our previous work\cite{maityStabilizingPolarDomains2023}. The smooth edition of the DP model \textit{se\_e2\_a} descriptor was used to train the MLFF using DeepMD-KIT\cite{zengDeePMDkitV2Software2023,zhangDeepPotentialMolecular2018}. The cutoff radius for nearest neighbors was set as 6 $\AA$ with a smoothing cutoff of 0.5 \AA. The model is comprised of an embedding net (10, 20, 40) and a fitting net (100, 100, 100) each with three layers. The architecture of the MLFF is described in detail in the original paper on DeepMD-Kit\cite{zengDeePMDkitV2Software2023}.

The training data includes the structures obtained from NPT MD simulations at the pressures 0.0, 0.7, 1.1, 1.5, 2.1, and 2.8 GPa obtained in a previous work conducted in our group~\cite{maityStabilizingPolarDomains2023}. This data was modified to correct for errors arising from fictitious kinetic energy associated with CPMD and further structures were included using concurrent learning using DP-GEN~\cite{zhangDPGENConcurrentLearning2020}. The details of these procedures and the subsequent training is described in detail in the SI. The final model obtained has a RMSE of energy, forces, and stresses as 0.42 meV/atom, 44.0 meV/$\AA$ and 0.02 GPa respectively.

\subsection{MD Simulations}

The MD simulations using the trained MLFF was carried out using LAMMPS~\cite{thompsonLAMMPSFlexibleSimulation2022}. A timestep of 0.5 fs was used for most simulations, although lesser timesteps of 0.2 or 0.1 fs were used in some cases. The MD simulations were carried out using the two cell sizes $4\times4\times4$ and $8\times8\times8$ which contains 768 and 6144 atoms respectively. The simulations were initialized from the cubic structure at ambient pressure in a $4\times4\times4$ cell. Simulations were carried out with this cell at the pressures 0.0, 0.4, 0.7, 1.1, 1.5, and 2.1 GPa with a timestep of 0.1 fs. A small timestep was used as we observed that larger timesteps sometimes led to instabilities in the MD. Starting from the cubic structure led to instabilities for an $8\times8\times8$ cell. Thus, the $8\times8\times8$ simulations were initialized by replicating a $4\times4\times4$ cell equilibrated  simulation at the same pressure. $8\times8\times8$ simulation cell with 6144 atoms. Equilibration was done for 150 ps and production runs were done for at least 250 ps at all pressures. For all the results, we use the $8\times8\times8$ supercell simulations, unless mentioned otherwise.

\section{Results and discussion}

\begin{figure*}[th]
	\centering
	\includegraphics[width=0.99\textwidth]{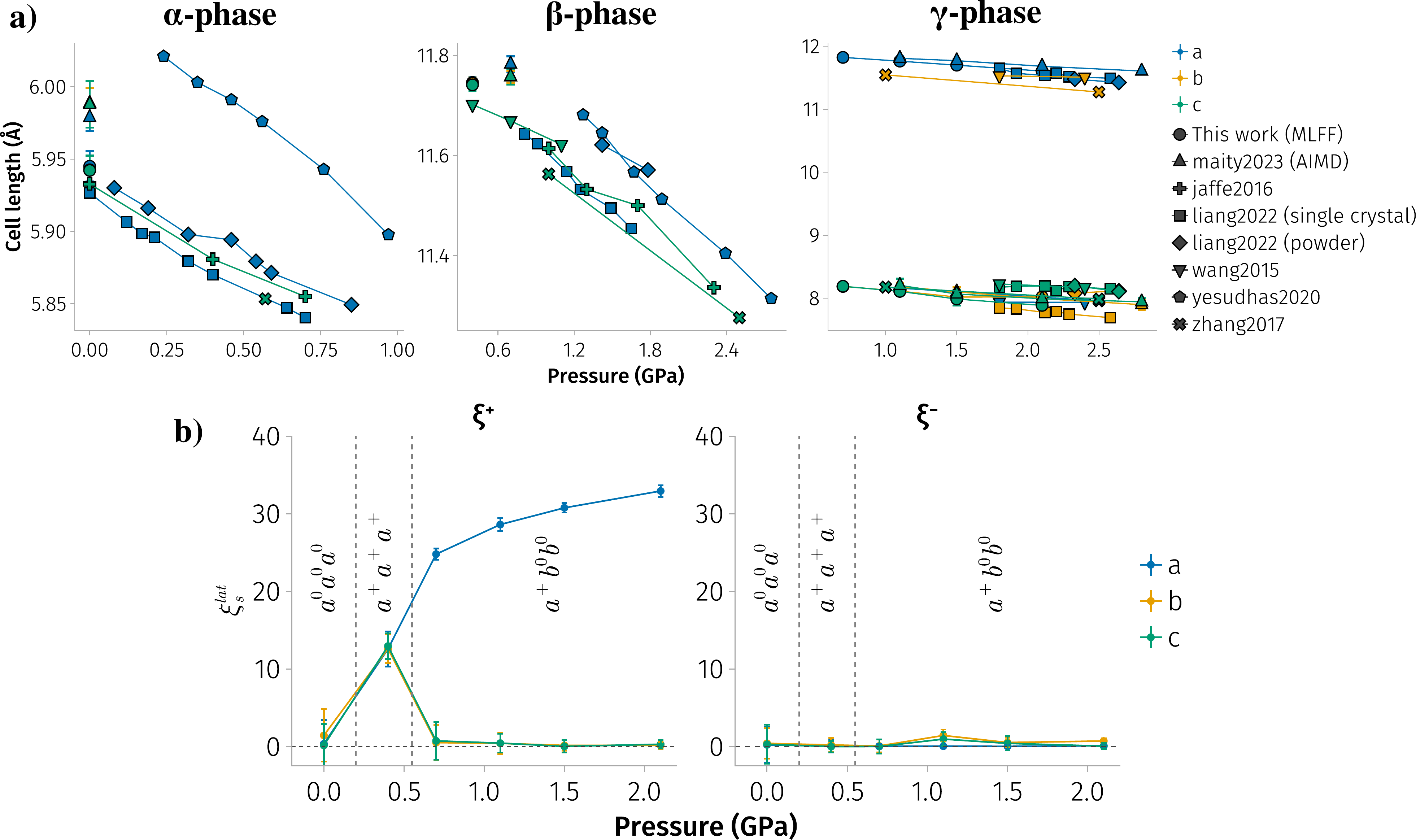}
	\caption{a) Comparison of cell parameters obtained from our MLFF MD simulations with previous experiments\cite{jaffeHighPressureSingleCrystalStructures2016,liangReassigningPressureInducedPhase2022,zhangEffectsNonhydrostaticStress2017,yesudhasCouplingOrganicCation2020} and simulations\cite{maityStabilizingPolarDomains2023}. The colors denote the directions \textit{a}, \textit{b} or \textit{c} and the symbols denote separate experiments/simulations. The \textit{b}-axis of the $Pmn2_1$ phase obtained in Ref.~\cite{liangReassigningPressureInducedPhase2022} has been halved from its original value for comparison purposes as a cell doubling occurs in this direction compared to the $Pnma$ phase reported by other studies. The value of lattice parameter $a$ in the $Pm\bar{3}m$ phase in Ref.~\cite{wangPressureInducedPhaseTransformation2015a} is very different from other works (8.44 $\AA$) and hence is not included. b) Average of the global order parameter associated with the lattice, $\xi_{s^\pm}^{lat}$ at various pressures, where $s = a, b$ or $c$ denotes the coupling direction.}
	\label{fig:panel1}
\end{figure*}

 \subsection{Global structural parameters}

The three pressure-induced phases of \ce{MAPbBr3}, denoted as $\alpha$, $\beta$ and $\gamma$, are obtained at the pressures 0.0 GPa, 0.4 GPa and 0.7 GPa respectively in our simulations. The average lattice parameters indicate that the structure remains cubic up to a pressure of 0.4 GPa. At higher pressures a tetragonal lattice is seen upto 1.5 GPa as indicated by fully overlapping distributions of the \textit{b} and \textit{c} lattice parameters. However, orthorhombic distortions start to appear as the pressure increases to 2.1 GPa and distinct distributions for \textit{b} and \textit{c} emerge. AIMD simulations indicated a tetragonal lattice at all pressures and orthorhombic distortions were not captured at higher pressures~\cite{maityStabilizingPolarDomains2023}, likely due to difficulty in resolving small differences of b and c in the small timescales sampled. A similar trend is observed in experiments as well where at lower pressures some experiments show very small differences between b and c. Hence we identify the phase observed from 0.7 GPa and above as the high-pressure $\gamma$ phase. All values obtained in the simulations are close to experimental values as shown in Fig.~\ref{fig:panel1} (a). In general, the phase transitions reported here occur close to the lowest experimentally reported pressures. These might likely arise due to pure hydrostatic conditions in simulations as compared to experiments where deviations from hydrostaticity might delay transition pressures.

\begin{figure*}[th]
	\centering
	\includegraphics[width=0.75\textwidth]{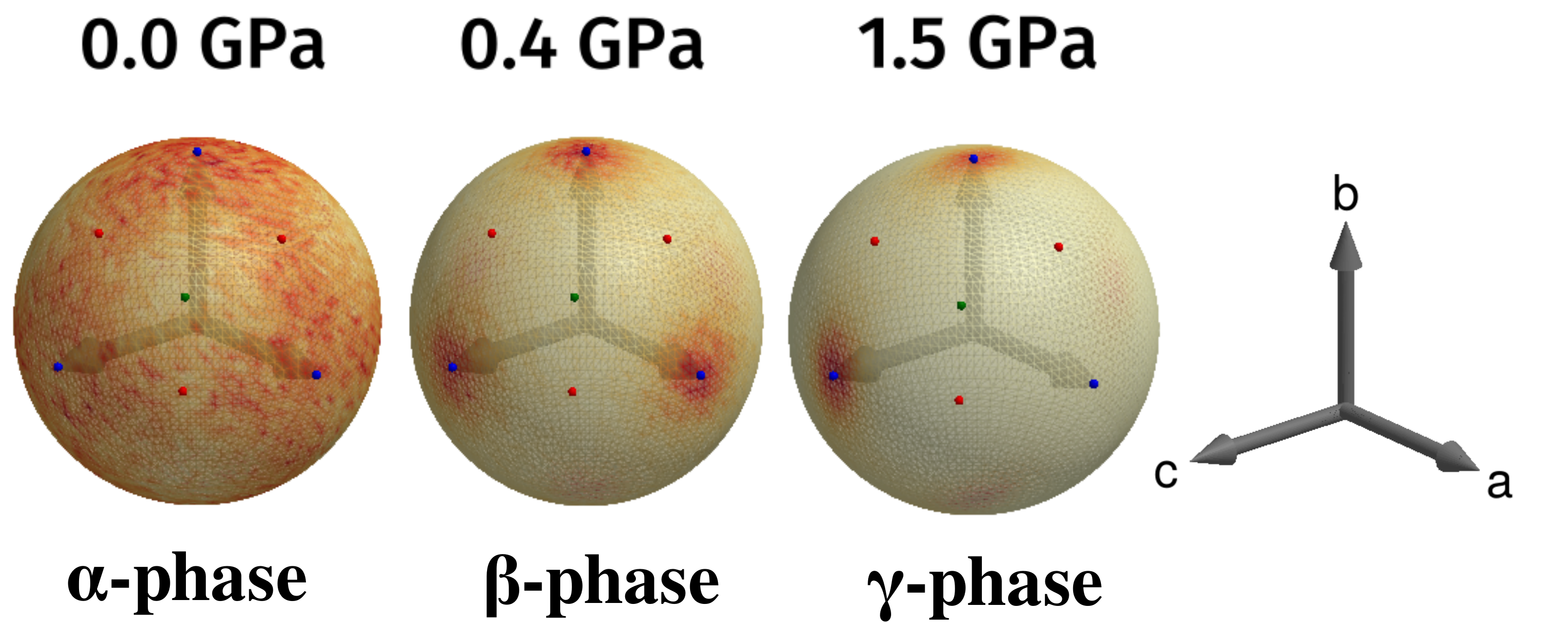}
	\caption{Histogram on a sphere surface depicting the orientational distribution of MA for each phase where each point represents a unique MA orientation ($\theta$, $\phi$). The arrows on the right shows the positive a, b and c axes directions to denote the regions in the histogram oriented towards the face centers along these directions. The high-symmetry orientations are marked with blue, green, and red circles denoting \textit{face-to-face}, \textit{body-diagonal}, and \textit{edge-diagonal} configurations, respectively. The negative directions as well as other pressures histograms are shown in the SI.}
	\label{fig:ssh_allphase}
\end{figure*}

The MA orientations and octahedral tilting behaviour of these phases can be probed using order parameters quantifying these motion for every unit cell and collecting these values for all unit cells while taking into account the phase change between neighboring unit cells. Such a procedure has been described previously~\cite{maityCooperativeOctahedralTilt2024a} and we use them here to quantify the octahedral (host) tilts as well as the MA (guest) orientations respectively using plane-projected order parameters denoted as $\xi_{s}^{lat}$ and $\xi_{s}^{MA}$, where \textit{s} represents the axis of tilt. The details of the order parameter computation is described in the SI. For octahedral tilting, in-phase and out-of-phase tilting between the layers are respectively distinguished as $\xi_{s+}^{lat}$ and $\xi_{s-}^{lat}$. In the glazer notation, these tilts are represented as $s^{0/+/-}$ where $s$ represents the direction and $0$, $+$ and $-$ respectively denote no-tilt, in-phase tilt and out-of-phase tilt. Hence, a non-zero value for $\xi_{a+}^{lat}$ and zero-value for all other tilt phases and directions will lead to a glazer notation of $a^+b^0b^0$. Equality of the tilt amounts for two directions are usually represented by the use of same symbol for the tilts, as done in this example. Note that a sign change of the value of the order parameter corresponds to a global phase change where all the tilts are reversed and hence only the magnitudes of these parameters are relevant. Using these order parameters, we obtain the MA orientational and octahedral tilting behavior for each of the three phases.

The average values of $\xi_{s\pm}^{lat}$ for each pressure simulated is shown in Fig.~\ref{fig:panel1} (b). In the $\alpha$ phase obtained at 0.0 GPa, the values are zero on an average for both in-phase and out-of-phase tilting indicating the absence of persistent octahedral tilts at this pressure. Hence a glazer notation of $a^0a^0a^0$ is assigned which corresponds to the tilting behavior observed experimentally in the $Pm\bar{3}m$ phase. As the pressure is increased to 0.4 GPa, a phase transition occurs to a structure with a non-zero in-phase tilt along all directions and zero out-of-phase tilts. As the amount of in-phase tilt along each direction are equal, it corresponds to a glazer notation of $a^+a^+a^+$ and represents the experimentally observed tilting behavior of the $Im\bar{3}$ phase. Although indications of such a tilt were present in previous AIMD simulations, the magnitudes were low with poor resolution~\cite{maityStabilizingPolarDomains2023,maityCooperativeOctahedralTilt2024a}. Such behaviour likely arises due to length-scale and time-scale issues, on which we discuss later, which is being resolved in our MLFF simulations. At higher pressures, a further transition is observed in the octahedral lattice tilts where a finite in-phase tilt is observed along one direction while very small out-of-phase tilts are present in the other two directions. As the value of out-of-phase tilts obtained is very close to zero, a glazer notation of $a^+b^0b^0$ is given. Out-of-phase tilts were absent in this phase for previous AIMD simulations as well. This tilting behavior is different from the experimentally obtained $a^+b^-b^-$ for the $Pnma/Pmn2_1$ phases.  We discuss on likely reasons for the same later.

The MA orientations measured via spherical polar coordinates $\theta$ and $\phi$ are represented using a histogram on a unit-sphere. Any point on its surface denotes a unique MA orientation. Such sphere-surface histograms representing each phase are shown in Fig.~\ref{fig:ssh_allphase}. Note that only the positive axes directions are shown here as no preference between any of positive/negative axes is present in our simulations. The corresponding 2-D distribution plots are shown in the SI. The MAs are orientationally disordered in the ambient pressure $\alpha$ phase sampling all the high-symmetry orientations more or less uniformly at 0.0 GPa, in agreement with previous AIMD~\cite{maityStabilizingPolarDomains2023}. In the $\beta$ phase, the majority of the distribution shifts to the face centers which were sampled uniformly in all directions, i.e, a partial ordering takes place. This uniformity between directions were absent in previous AIMD simulations where a slight preference for one of the planes was seen for the MAs\cite{maityStabilizingPolarDomains2023}. This occurs due to unconverged sampling of MA reorientational distribution in AIMD due to the smaller simulation time scales. This distribution also agrees with experimentally reported structure of MAs in the $Im\bar{3}$ phase, where three fourths of the MAs points towards a face\cite{jaffeHighPressureSingleCrystalStructures2016}. At even higher pressures, starting from 0.7 GPa in the $\gamma$ phase, the MAs gets restricted in a plane where the 4 faces in the \textit{b-c} plane are sampled as seen in prior AIMD. Additionally, we also observe long-range staggered ordering (LRSO) in this phase, reported to occur in previous AIMD simulations\cite{maityStabilizingPolarDomains2023} which we discuss in detail later. The ratio of occurrence (RO) of the high-symmetry orientations shows a clear preference for the edge-diagonal orientation is present at ambient pressure, whereas, at all higher pressures the face-to-face orientation is preferred and its likelihood increases with pressure. Such a behaviour was observed in previous AIMD simulations as well, although the ratios obtained here have changed  due to improved sampling\cite{maityStabilizingPolarDomains2023}. We further obtain the MA reorientation timescales in all the phases and is shown in the SI. The obtained timescales are slightly overestimated than the corresponding timescales for AIMD\cite{maityStabilizingPolarDomains2023}. This likely occurs as a result of slight volume underestimation with respect to AIMD, which will increase the reorientation barriers and hence increase reorientation times. 

Next, we discuss in detail the dynamics of host and guest in each of the phases obtained and answer relevant questions regarding each of them.

\subsection{Low pressure $\alpha$ phase}

The ambient pressure $\alpha$ phase with $Pm\bar{3}m$ space group is characterized mainly by its completely disordered MA orientations and absence of non-zero persistent tilt modes. Although this phase has cubic symmetry, it has been claimed to display local distortions.  There have been some claim of persistent distortions in the ambient pressure cubic phase\cite{pageShortRangeOrderMethylammonium2016,bernasconiDirectEvidencePermanent2017}. The distribution of local lattice parameters we obtain do not indicate the presence of such distortions, agreeing with previous AIMD results~\cite{maity_deciphering_2022}. Recently it has also been found that this phase accommodates dynamic twinned nanodomains exhibiting finite out-of-phase octahedral tilts of tetragonal symmetry\cite{dubajic_dynamic_2025}. However, the cell sizes we presently utilize are insufficient to capture such nanodomains, which may be at least 6 unitcells wide.

Previous AIMD simulations have shown that a triple-well forms for the octahedral tilt modes in this phase due to stabilization of non-zero amplitudes of tilt. Such a triple-well formation was not clear in the distributions of octahedral tilt amplitudes obtained in this phase. However, distribution of layerwise tilt order parameter shows a flat-topped peak with a broad range of values sampled. Such a flat peak distribution likely arises from the triple-well character of the underlying potential energy surface. Decomposition of the layerwise distribution further into distributions for individual layers, clearly shows that a triple-well behaviour is displayed by most layers. This was confirmed by fitting these layerwise distributions to a 3-component Gaussian Mixture Model (GMM) which showed that two of the three peaks have finite non-zero values ($\sim \pm$9.0$^{\circ}$) and the third peak is at zero. It can also be seen that the flat-peaked nature of the combined distribution over all the layers arises from the triple-well nature of the individual layers which as depicted by the GMM fit of the combined distribution. Although, the large deviations from the peak positions (average covariance of the components of the GMM is 16.5$^\circ$) in comparison to the closeness of the peaks, modulates the central peak intensity compared to other peaks, making the triple well character less visible. Thus, our simulations reproduce the previously predicted triple well character of the octahedral tilts in the $\alpha$ phase .

\subsection{Intermediate pressure $\beta$ phase}

\begin{figure*}
	\centering
	\includegraphics[width=0.99\textwidth]{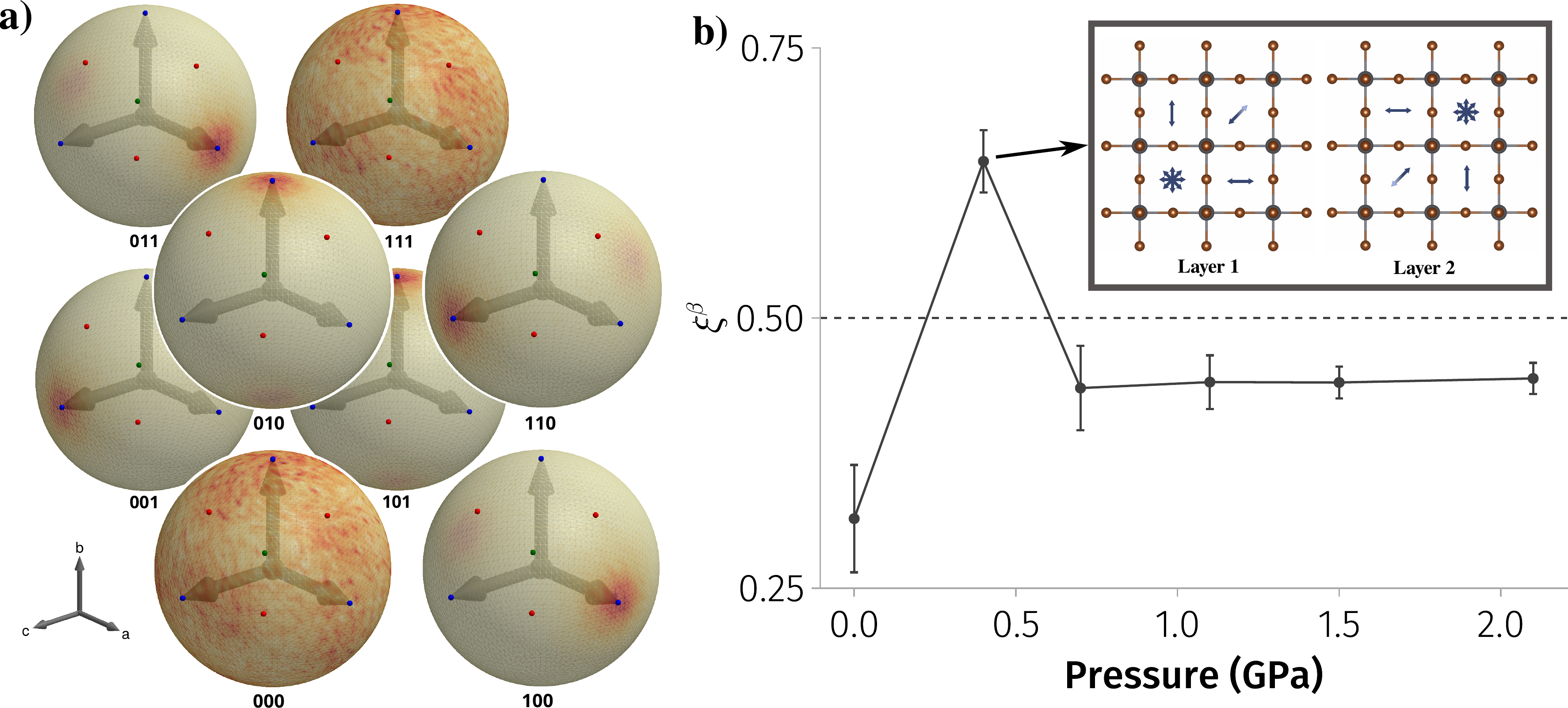}
	\caption{Intermediate pressure MA ordering. a) Orientational distribution of each MA in the $2\times2\times2$ supercell formed after cell doubling. b) Ensemble averages of the intermediate pressure global order parameter, $\xi^{\beta}$ corresponding to the origin with the maximum value. Inset: Schematic representing the orientational ordering of each of the MA.} 
	\label{fig:panel2}
\end{figure*}

The intermediate pressure $\beta$ phase displays equal in-phase tilt along all directions and the MA orientations appears to be disordered over all the 6 faces (Fig.~\ref{fig:panel1} (a), (b)). However, previous experiment report indicates that two kinds of MAs are present in this phase distinguished by the type of disorder each MA possess\cite{jaffeHighPressureSingleCrystalStructures2016}. However, such a behavior was not observed in previous AIMD simulations, likely due to the smaller time scales\cite{maityStabilizingPolarDomains2023}. Here, we look into whether such a behavior is present by monitoring the MA distributions of individual unit cells from our simulations. Although the global orientational distribution shows sampling of all the face centers homogeneously, individual distributions give a completely different picture. These distributions show that each MA can be classified into two classes based on its orientational distribution and these two classes arrange in a pattern leading to the formation of a $2\times2\times2$ MA sublattice comprising of 8 pseudo-cubic unit cells (Fig.~\ref{fig:panel2} (a)). The first class appears to behave orientationally disordered with a distribution covering all the orientations more or less uniformly. In the second class, the molecular axis of MAs are ordered and orient towards one of the faces along any of the three directions. Hence, the MA sublattice contains 2 MAs which are orientationally disordered and 2 MAs oriented towards each of the faces along the three directions forming an 8-membered $2\times2\times2$ cell. Further, these 8 MAs are arranged as two 4-membered layers each containing one disordered MA and the other three oriented perpendicular to each of the three unique faces. These layers are connected to each other by inversion symmetry as shown in Fig.~\ref{fig:panel2} (b) inset. A calculation of the reorientation times of these two classes of MAs gave a fast reorientation time of 4.5 ps for the disordered MA whereas the timescale for the ordered MA was obtained to be $\sim$20 ps, indicating the rapid motion of the disordered MA. This classification of MAs is in agreement with the aforementioned experimental study which observed two inequivalent A-site cations with different type of disorder and matches the structure they report~\cite{jaffeHighPressureSingleCrystalStructures2016}. The experimental study hypothesized that the formation of these two type of A-site cations may originate as a result of different H-bonding interactions within the two sites. However, the origin of different interactions in the two sites is unclear. Here, we first quantify this interesting MA ordering using an order parameter which picks up the presence of such ordering from the simulations. Subsequently, the origin of this ordering is investigated based on the octahedral tilt behavior present in this phase.

To quantify the ordering present in this phase, we define an order parameter that essentially captures the alignment of MA molecules with respect to the ideal ordering described above, similar to previously defined order parameters capturing the long-range staggered order~\cite{maityCooperativeOctahedralTilt2024a}. To achieve the same, the MA orientation described via $\theta$ and $\phi$ are used to define local 3-D order parameters as,
$$
\vec{\eta}^{\beta}_{\vec{l}} = \left[ \cos\phi \sin\theta, \sin\phi \sin\theta, \cos\theta \right]
$$
These are essentially the cartesian coordinates of the unit vector representing the MA orientation, $(\theta, \phi)$. It is to be noted that, for the ordered MAs, only the MA molecular axis is ordered and the MA dipoles remain disordered for all cases. Hence, the ranges of $\theta$ and $\phi$ were modified so as to make the two dipoles of a particular molecular axis equivalent as shown in the SI. For constructing the order parameter, we align the molecular axis of all MAs along the \textit{z}-axis by rotating the corresponding $\vec{\eta}^{\beta}_{\vec{l}}$ based on the index of the unit cell, $\vec{l} = (l_x, l_y, l_z)$. 
%As a $2\times2\times2$ MA sublattice is formed, we take a modulus by 2 of $\vec{l}$, $\vec{l}_{mod}$ as $(l_x \text{ mod } 2, l_y \text{ mod } 2, l_z \text{ mod } 2)$. Subsequently the MAs are aligned along the \textit{z}-axis, by application of rotation matrices, $U(\vec{l})$ defined as,
This was done by application of the rotation matrices, $U(\vec{l})$ defined as,
\[
U(\vec{l}) = 
\begin{cases}
	I_3, & \text{if } \vec{l} = (0,0,0) \text{ or } (1,1,1) \\[10pt]
	I_3, & \text{if } \vec{l} = (1,0,0) \text{ or } (0,1,1) \\[10pt]
	R_y\left( -\frac{\pi}{2} \right), & \text{if } \vec{l} = (0,1,0) \text{ or } (1,0,1) \\[10pt]
	R_x\left( \frac{\pi}{2} \right), & \text{if } \vec{l} = (0,0,1) \text{ or } (1,1,0)
\end{cases}
\]
where $I_3$ is the $3 \times 3$ the identity matrix and $R_y$, $R_x$ represents rotation about the \textit{y} and \textit{x} axes respectively. Note that no rotation is done for the first case as the corresponding MA is disordered, while for the second case the corresponding MA is assumed to be already aligned along the \textit{z}-axis (Fig.~\ref{fig:panel2} (a)). We note that any of the four unit cells shown in Fig.~\ref{fig:panel2} (a) may be aligned along the \textit{z}-axis. This corresponds to a shift of the origin corresponding to a global phase for the order parameter. Hence, there are four possible phases (origins) with which this order parameter can be computed for a simulation and these correspond to various permutations of $U(\vec{l})$. Thus, the rotation matrices defined by $U(\vec{l})$ is used to rotate $\vec{\eta}^{\beta}_{\vec{l}}$ to align them along the z-axis as,
$$
\vec{\omega}^{\beta}_{\vec{l}} = U(\vec{l}) \cdot \eta^{\beta}_{\vec{l}}
$$
A global vector order parameter can be obtained by taking the average of $\vec{\omega}^{\beta}_{\vec{l}}$ over the lattice to get $\vec{\Omega}^{\beta}$.
The projection of this vector along the z-axis gives the global order parameter $\xi^{\beta}$ as,
$$\xi^{\beta} = [0,0,1] \cdot \vec{\Omega}^{\beta}$$
As one is free to choose any of the four global phases while computing $\xi^{\beta}$, we have computed the ensemble average of $\xi^{\beta}$ for all four possible phases as shown in the SI. In the $\beta$ phase at 0.4 GPa, $\xi^{\beta}$ takes a large value of $\sim$0.67 for the correct origin, while for other origins $\xi^{\beta}$ displays smaller values. The value is not equal to 1.0 as the disordered MA is unaligned most of the time and reduces the $\xi^{\beta}$ values. It can be seen that for the $\alpha$ phase at 0.0 GPa, $\xi^{\beta}$ takes a constant value of $\sim$0.3 for all four origins, as each MA can be expected to face a particular axis about one third of the time due to its disordered nature. For the $\gamma$ phase at pressures above 0.7 GPa, two possible values, one close to 0.4 and the other close to 0.0, are sampled for various origins. This occurs as LRSO of MAs in a layer, present in the $\gamma$ phase, is equivalent to the ordering in the $\beta$ phase for half the MAs for some origin choices while for the other origins their values cancel each other. These show that the maximum value of $\xi^{\beta}$ out of the four choices will correspond to the correct origin as this would align the most with the MA ordering of the $\beta$ phase. Hence, the ensemble average of $\xi^{\beta}$ corresponding to the maximum value out the four origins are shown in Fig.~\ref{fig:panel2} (b) for all the pressures. It can be clearly seen that at 0.4 GPa in the $\beta$ phase, this value becomes high ($> 0.5)$, while for all other pressures a low value ($< 0.5$) is observed. Thus, this order parameter captures the MA sublattice doubling present in this phase.

\begin{figure}
	\centering
	\includegraphics[width=0.3\textwidth]{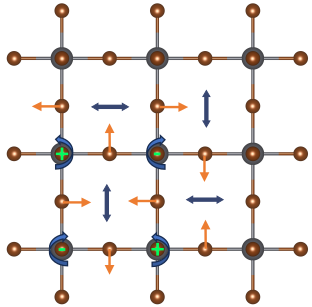}
	\caption{Preference for MA alignment towards horizontal and vertical axis arising as a result of anti-clockwise (positive) and clock-wise (negative) tilts respectively.} 
	\label{fig:tilt_MA_preference}
\end{figure}

The MA sublattice doubling may originate as a result of the global in-phase tilt ordering of this phase represented by the glazer notation $a^+a^+a^+$. However, how the tilts leads to such a behavior has not been addressed yet. It has been previously pointed out that, an in-phase tilt along a particular axis will lead to the alignment of MAs along one of the two axis in the plane perpendicular to the tilt axis, in the high pressure $\gamma$ phase\cite{maityStabilizingPolarDomains2023}. Additionally, a staggered ordering (LRSO) emerges where neighboring unitcells prefer alignment along perpendicular axes. This is a result of strong guest-host coupling arising from optimal H-bonding interactions with the Br along one axis in the plane. Also, a reduction in volume along the other axis due to the tilting leads to non-optimal H-bonding interactions along this axis. These effects forces the MA to align along the elongated axis as shown in Fig.~\ref{fig:tilt_MA_preference}. Hence, in a single unitcell, the an anti-clockwise tilt (positive tilt angle) of an octahedra will force the MA in the top-right of this octahedra towards the horizontal axis, whereas a clockwise tilt (negative tilt angle) will force it towards the vertical axis. However, in contrast to the $\gamma$ phase where in-phase tilt is present along just one axis, here in-phase tilt is present along all three axes.  Thus, tilts along each of the three axes may lead to different axes of optimal H-bonding interactions and their combined effect would determine the final MA alignment. One can compute a measure of total preference towards each axis by assigning a +1 value for an increased preference and -1 value for a reduced preference. These are shown in Table~\ref{tab:tilt_maaxis_preferences} for various combination of tilts.

\begin{table}[h]
	\centering
	\renewcommand{\arraystretch}{0.95} % Reduces row height
	\begin{tabular}{|c|c|c|c|c|}
		\hline
		Tilt-direction & Tilt-sign & \multicolumn{3}{c|}{MA axis preference} \\
		\cline{3-5}
		& & a & b & c \\
		\hline
		a & + & 0 & +1 & -1 \\
		%		\hline
		a & - & 0 & -1 & +1 \\
		%		\hline
		b & + & -1 & 0 & +1 \\
		%		\hline
		b & - & +1 & 0 & -1 \\
		%		\hline
		c & + & +1 & -1 & 0 \\
		%		\hline
		c & - & -1 & +1 & 0 \\
		\hline
	\end{tabular}
	\caption{MA axis preferences for different kind of tilts}
	\label{tab:tilt_maaxis_preferences}
\end{table}

\begin{figure*}
	\centering
	\includegraphics[width=0.95\textwidth]{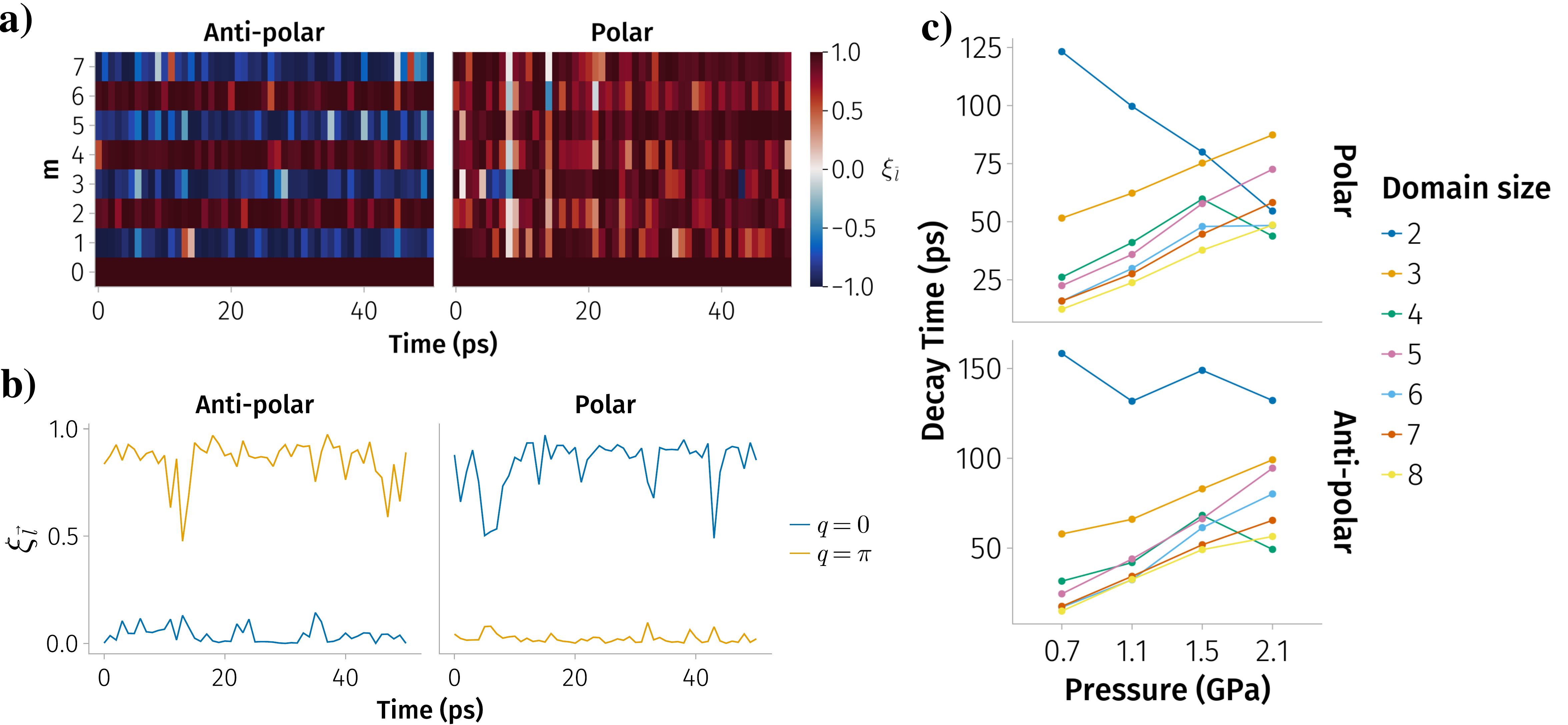}
	\caption{Distributions of tilt patterns (a) and the associated MA orientation distributions (b) for the four groups of unitcells in the simulation cell displaying a distinct orientational MA alignment. Here the tilt patterns were obtained by normalizing a vector composed of the tilts along the three directions as $\vec{K}=[\eta_a^{lat}, \eta_b^{lat}, \eta_c^{lat}]$ and converting them to polar coordinates, ($\theta$, $\phi$). Each tilt component $\eta_s^{lat}$ is obtained as the magnitude of unitcell lattice tilt order parameter $\vec{\eta}_s^{lat}$} 
	\label{fig:tilt_MA_distribution}
\end{figure*}

For a single unitcell, these preferences will add up for the tilt along each axis to give a total preference for the alignment of MA towards each axis in the unitcell. The two possible signs of tilts along each of the three tilt directions results in 8 unique combinations of tilts which a unitcell can possess. Computing the total MA alignment preference for each of these 8 possible tilts will hence give the possible MA orientations. These are shown in Table~\ref{tab:tilt_maaxis_orientation}. It can be seen that two kinds of MA orientations arise from these 8 possible tilt patterns. These combination of tilts can either lead to a situation where none of the axes are preferred resulting in disorder or a high preference for one of the axis is seen resulting in alignment towards it. Thus, the combination of tilts along the three directions effectively leads to the decomposition of MAs into two kinds, one with disordered MA molecular axis and the other with ordered molecular axis. In our simulations, we see that the unitcells only sample 4 out of the 8 possible tilt patterns listed in Table~\ref{tab:tilt_maaxis_orientation}. These 4 tilt patterns were indeed associated with the corresponding MA orientational distributions as depicted in Fig.~\ref{fig:tilt_MA_distribution}.

\begin{table}[h]
	\centering
	\renewcommand{\arraystretch}{0.95} % Reduces row height
	\begin{tabular}{|c|c|c|c|c|c|c|}
		\hline
		\multicolumn{3}{|c|}{Tilt-sign} & \multicolumn{3}{|c|}{\shortstack{MA axis\\preference}} & Final preference \\
		\cline{1-6}
		a & b & c & a & b & c & \\
		\hline
		+ & + & + & 0 & 0 & 0 & Disorder\\
		+ & + & - & -2 & +2 & 0 & b \\
		+ & - & + & +2 & 0 & -2 & a \\
		+ & - & - & 0 & +2 & -2 & b \\
		- & + & + & 0 & -2 & +2 & c \\
		- & + & - & -2 & 0 & +2 & c \\
		- & - & + & +2 & -2 & 0 & a \\
		- & - & - & 0 & 0 & 0 & Disorder \\
		\hline
	\end{tabular}
	\caption{MA axis preferences for each combination of tilts}
	\label{tab:tilt_maaxis_orientation}
\end{table}

As we had pointed out earlier, the aforementioned behavior of MA orientations were not captured in AIMD simulations in a smaller $4\times4\times4$ cell~\cite{maityStabilizingPolarDomains2023}. In addition, our MLFF simulations with a large timescale did not capture the same as well. Although the AIMD tilt showed indications of in-phase tilt behavior along all directions, the corresponding values were small and unequal. We look into the tilt behavior displayed in our $4\times4\times4$ supercell simulations to investigate into possible reasons for the same. Here it is seen that the although the global order parameter displays finite non-zero tilts along all directions, a disorder exists for the tilt directions. This indicates existence of layerwise tilt disorder where each layer samples both tilt directions. Such a layerwise tilt disorder was observed in the AIMD simulations as well\cite{maityCooperativeOctahedralTilt2024a}. However, the larger $8\times8\times8$ cell simulations do not display such a tilt disorder and only a single tilt direction is sampled for a particular tilt axis. Thus, the layerwise tilt disorder observed in the smaller cell simulations indicates that the barrier for changing tilt direction is smaller in this cell size in comparison with the larger cell. These could arise due to large fluctuations in a small scale and the ease of it extending throughout the cell in a smaller cell, while extension of the same throughout the cell in larger cell will be hindered. Presence of the aforementioned layerwise tilt disorder in smaller cell hinders the MA sublattice doubling. Thus, the length-scale of the simulation plays an important role in stabilizing some features of this phase.

\subsection{High pressure $\gamma$ phase}

The octahedral tilt behavior determined for the high pressure $\gamma$ was $a^+b^0b^0$, which is different from the the experimentally predicted $a^+b^-b^-$. The obtained out-of-phase tilts were very small to characterize them as out-of-phase (Fig.~\ref{fig:panel1} (b)). Also, persistent out-of-phase tilts require has been associated with permanent off-center MA displacements which couple to the Br, based on results from the low temperature \textit{Pnma} phase\cite{maityCooperativeOctahedralTilt2024a,maity_deciphering_2022}. Such off-center displacements were not observed at any of the pressures indicating the absence of permanent out-of-phase tilts. As described earlier, the orthorhombic character of this phase was only visible at high pressures of 2.1 GPa, likely since the difference between \textit{b} and \textit{c} at lower pressures is very small, as indicated by some experiments. Thus, our simulations up to pressures of 1.5 GPa have approximate tetragonal symmetry, which does not accomodate out-of-phase tilts in combination with in-phase tilts. Hence, it is likely that the approximate tetragonal character is leading to very small out-of-phase tilts which can't be distinguished from fluctuations. However, we do see indications of such out-of-phase tilt character from the individual layerwise tilt distributions. Here, along the \textit{b}-direction it can be seen that although each layer has distributions peaked at off-zero values, and every consecutive layers has a peak with an opposite sign to the previous layer, denoting out-of-phase tilt. However, these are not obtained for all directions or pressures likely due to reasons mentioned above.

\begin{figure*}[th]
	\centering
	\begin{subfigure}[t]{0.49\textwidth}
		\centering
		\includegraphics[width=\linewidth]{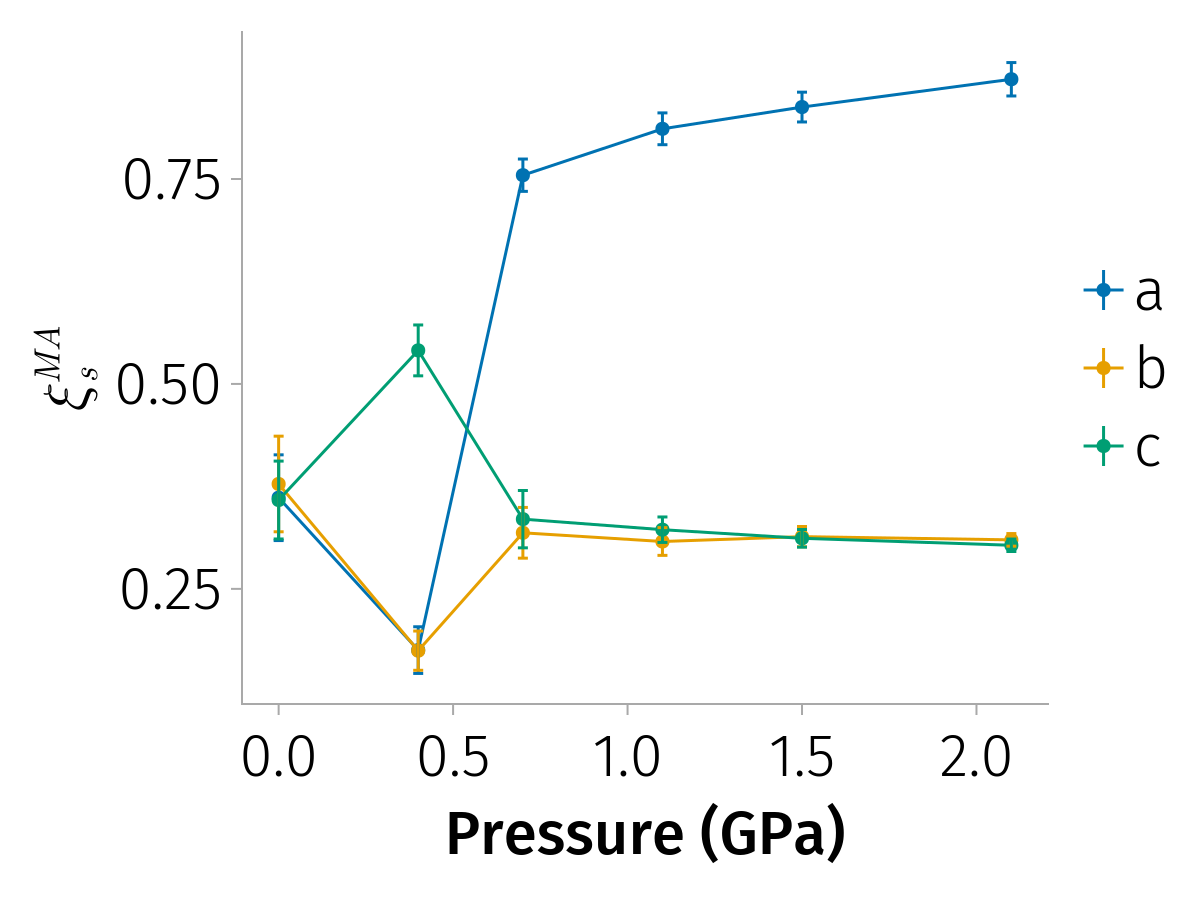}
		\caption{}
	\end{subfigure}\hfill
	\begin{subfigure}[t]{0.49\textwidth}
		\centering
		\includegraphics[width=\linewidth]{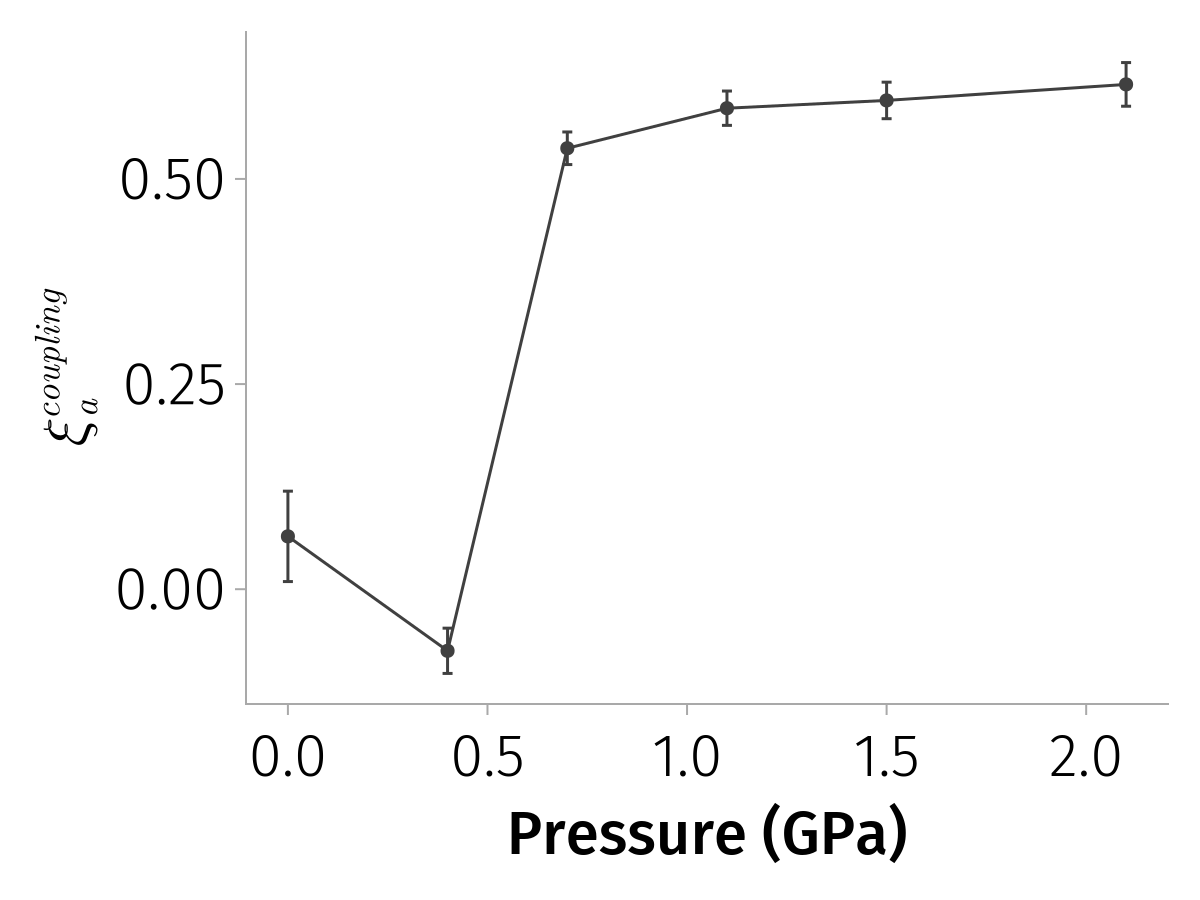}
		\caption{}
	\end{subfigure}\hfill
	\caption{(a) Average of the global order parameter associated with MA, $\xi_{s^\pm}^G$ at various pressures, where $s = a, b$ or $c$ denotes the coupling direction. (b) Average of the global order parameter associated with lattice-MA coupling along the unique-axis \textit{a}, $\xi_{a}^{coupling}$ at various pressures.}
	\label{fig:xi_ma_coupling_avg}
\end{figure*}

The high pressure $\gamma$ phase is also characterised by long-range staggered ordering (LRSO) of the MA orientations. The LRSO of MA orientations have been quantified previously \cite{maityCooperativeOctahedralTilt2024a} using guest order parameter $\xi_s^{MA}(\vec{q},t)$ where $\vec{q}= \frac{2\pi}{a}\left(0,\frac{1}{2},\frac{1}{2}\right)$ where $s$ denotes the coupling direction. We compute the same here as well for the phases simulated as shown in Fig.~\ref{fig:xi_ma_coupling_avg} (a). It can be seen that, in the $\gamma$ phase starting at 0.7 GPa, these order parameters increase sharply denoting LRSO. Interestingly, intermediate values are seen for the $\beta$ phase at 0.4 GPa, likely due to the ordering of a majority of MAs along face-to-face orientations in this phase. It has also been shown that the coupling between MA orientations and octahedral tilt modes plays an important role in restricting the MA dipoles in a plane in the high-pressure phase\cite{maityStabilizingPolarDomains2023,maityCooperativeOctahedralTilt2024a}, which were quantified using the coupling order parameter $\xi_s^{coupling}$. These indicate large coupling present in the $\gamma$ phase at all pressures, whereas the coupling strength indicated by this parameter is significantly low for other phases (Fig.~\ref{fig:xi_ma_coupling_avg} (b)), emphasizing the role of host-guest coupling towards the restriction of the MA in the plane. The long-range orientational order present in this phase can be compared to liquid crystals and has been described previously using nematic order parameters\cite{maityStabilizingPolarDomains2023}. The transition to an oblate nematic phase is reproduced here, in this phase as well using the nematic order parameters.

Although staggered ordering leads to perpendicular MA orientations between neighboring unitcells in the same layer, the MAs in the neighboring unitcells in the different layers are oriented along the same axis. This leads to the formation of \textit{polar} or \textit{anti-polar} domains when the MA molecule in the neighboring layer is parallel or anti-parallel respectively. The existence of such domains can contribute to increase in carrier recombination rates.\cite{qiaoFerroelectricPolarizationSuppresses2019} In previous AIMD simulations, this emerged as an increased tendency to form polar and anti-polar domains in the $\gamma$ phase\cite{maityStabilizingPolarDomains2023}. However, due to the limited timescale and length scale of the simulations, the analysis on the dynamics of such domains were limited. Here, we reproduce the increased tendency to form polar and anti-polar domains in this phase. The nearest neighbor MA dihedral angles defined as, $\theta^{\rho}_{NN} = \phi^{\mu\nu}_2-\phi^{\mu\nu}_1$ where $\phi^{\mu\nu}$ are the azimuthal angles in the plane perpendicular to the $\rho$ direction along which pairs of MA neighbors ($\phi_1,\phi_2$) were computed. In the $\gamma$ phase, $\theta^a_{NN}$ assumes either $0^\circ$ (polar) or $\pm 180^\circ$ (anti-polar) indicating the formation polar or anti-polar domains in this phase. 

To monitor the dynamics of such domains, we define a complex order parameter to capture their formation along each chain. For every member, $m$ of a chain indexed by $\vec{l}$, the dihedral angle with respect to the first MA in the chain, $\theta_{\vec{l},m}$ is used to define an order parameter as,
$$
Z_{\vec{l},m}(t) = \cos \theta_{\vec{l},m}(t) + i \sin \theta_{\vec{l},m}(t) = e^{i\theta_{\vec{l},m}}
$$
where $i = \sqrt{-1}$ is the imaginary unit. The neighbors are along the unique axis of the $\gamma$ phase. These individual order parameters are \textit{collected} for each chain after performing a $q$-dependent \textit{local staggering} to obtain $Z_{\vec{l}}(q,t)$ as,
$$
Z_{\vec{l}}(q,t) = \frac{1}{N} \sum_{m=0}^{N-1} e^{-iqm}Z_{\vec{l},m}(t)
$$
where N is the number of units in the chain. The \textit{local staggering} is performed using a rotation in the complex plane defined as $e^{-iqm}$. Here, $q=0$ denotes a polar domain and $q=\pi$ denotes an anti-polar domain. This \textit{collection} leads to an average over all the units in a chain for the polar domain whereas neighboring units are rotated by $180^\circ$ before taking an average for an anti-polar domain. The expression of $Z_{\vec{l}}(q,t)$ is essentially a discrete Fourier transform and we are interested in the Fourier coefficients corresponding to a pattern which repeats every unit ($q=0$, polar) or repeats every two units ($q=\pi$, anti-polar). These coeffiecients are therefore complex numbers for which the associated magnitudes are obtained as,
$$
\xi_{\vec{l}}(q,t) = Re\left\lbrace Z_{\vec{l},0}^*(t) Z_{\vec{l}}(q,t) \right\rbrace
$$
The value of $\xi_{\vec{l}}(q,t)$ in turn tells how close a chain is to a polar ($q=0$) or anti-polar ($q=\pi$) structure. An example of the same is shown in the SI for a polar ($Pmn2_1$) and an anti-polar ($Pnma$) structure, wherein the value becomes 1.0 for $\xi_{\vec{l}}(q=0)$ and $\xi_{\vec{l}}(q=\pi)$ respectively for all the chains. The temporal changes in such domains can be visualized by looking at the magnitude of each unit cell obtained after projecting them onto the first unit as,
$$
\xi_{\vec{l},m}(t) = Re\left\lbrace Z_{\vec{l},0}^*(t) Z_{\vec{l},m}(t) \right\rbrace
$$

\begin{figure*}
	\centering
	\includegraphics[width=1.0\textwidth]{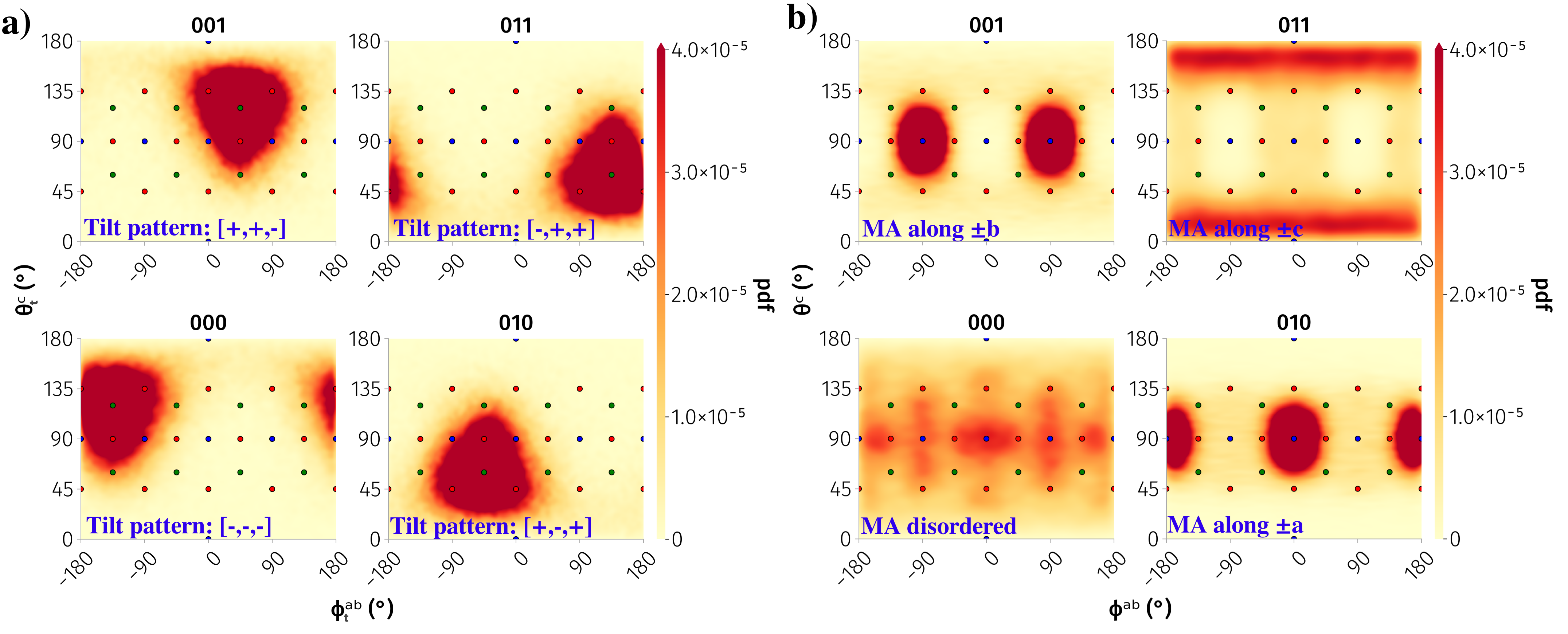}
	\caption{a) Polar domain order parameter for individual unit cells, $\xi_{\vec{l},m}(t)$ for an example anti-polar and polar chain for a 50 ps time frame b) Polar domain order parameter, $\xi_{\vec{l}}(q,t)$ for both the corresponding chains c) Decay times of polar ($q=0$) and anti-polar ($q=\pi$) domains for different domain sizes as a function of pressure. }
	\label{fig:panel4}
\end{figure*}

These order parameters can also be used to look into the behavior of domains with different sizes. The chain order parameters $Z_{\vec{l}}(q,t)$ and $\xi_{\vec{l}}(q,t)$ denotes domains with a size corresponding to the length of the chain, $N$. It is possible to define domains with size $M \leq N$ where $M$ denotes the size of the chain as,
\begin{align*}
	Z_{\vec{l}}^{(M)}(q,t) &= \frac{1}{N} \sum_{m=0}^{M-1} e^{-iqm}Z_{\vec{l},m}(t) \\
	\xi_{\vec{l}}^{(M)}(q,t) &= Re\left\lbrace Z_{\vec{l},0}^*(t) Z_{\vec{l}}^{(M)}(q,t) \right\rbrace
\end{align*}
where $2 \leq M \leq N$. For any chain with length $N$, there are $N$ ways to achieve a domain of size $M$, as there are $N$ possible starting units ($m=0$) and each will lead to to a different value of $\xi_{\vec{l}}^{(M)}(q,t)$. The dynamics of these domains can hence be quantified using autocorrelations of this order parameter defined as,
$$
C^{(M)}_{\xi}(q,\tau) = \frac{\left\langle \left\langle \xi_{\vec{l}}^{(M)}(q,t)\cdot \xi_{\vec{l}}^{(M)}(q,t+\tau) \right\rangle \right\rangle}{\left\langle \left\langle \xi_{\vec{l}}^{(M)}(q,t) \cdot \xi_{\vec{l}}^{(M)}(q,t) \right\rangle \right\rangle}
$$
where $\left\langle \left\langle ... \right\rangle \right\rangle$ denotes an ensemble average over all possible chains with time. These correlation functions for each domain size $M$ were fitted to stretched exponentials $e^{-(t/\tau)^\beta}$, typically used for disordered systems, to determine the decay time $\tau$. 

The timescales thus obtained for each domain size for both polar ($q=0$) and anti-polar ($q=\pi$) domains are shown in Fig.~\ref{fig:panel4} (c). As expected, the anti-polar domains have higher lifetimes than the corresponding polar ones. Also, larger domains have smaller lifetimes, although such domains persist for long times at higher pressures. For instance, at 1.5 GPa and higher pressures, domains of size 4 and higher have lifetimes exceeding 50 ps. Interestingly, at lower pressures, domains of size 2 have large lifetimes which decreases with pressure. This increase is significant for the case of polar domains, compared to that for anti-polar domains. This indicates polar minima gets stabilized locally at lower pressures, while this effect reduces as the pressure increases. In general, it can be seen that both polar and anti-polar domains persist for times upto 75 ps, which will have significant role in photovoltaic properties by increasing the carrier recombination rates. 

We also compute the persistence times of these domains defined as the average time a domain of a particular size persists, before a change occurs. These were computed by discretising the dihedral angles between the first MA and all other MAs in a chain into either +1 (parallel) or -1 (anti-parallel). These discretisation allows us to monitor polar and anti-polar domains at each frame determined as repetitions of the same value. The domains obtained are monitored for changes. The time-dependent evolution of domains of various sizes existing in the simulation are obtained in this manner which can be used to compute the average time a domain of a particular time persists in the simulation. The obtained persistence time for each domain sizes for all the pressures in the $\gamma$ phase are shown in the SI. The values obtained are significantly lower than the ones obtained previously, for the decay times, as one of the MAs reorienting in a single frame is enough to break the domain. The obtained persistence times at 2.1 GPa are $\sim$ 3 ps for a size 2 anti-polar domain and $\sim$ 1 ps for a size 6 anti-polar domain. The previous AIMD simulations showed slightly higher value of $\sim$ 5 ps for the size 2 domains at the pressure of 2.1 GPa, slightly higher than the lifetime obtained here\cite{maityStabilizingPolarDomains2023}. The reduction might possibly arise due to the large sampling obtained here improving the statistics to compute the same. Additionally, the AIMD simulations could not monitor larger size domains due to the limited length scales. Here we monitor the persistence times of domains with sizes upto 7.

\section{Conclusions}

We have trained a MLFF for the pressure-induced phases of \ce{MAPbBr3} capable of accurately representing the $\alpha$, $\beta$ and $\gamma$ phases. This MLFF was used to conduct a detailed investigation into these phases using simulations long time scales and large length scales. The properties of these phases were analyzed in terms of the behavior of host octahedral framework and guest MA molecules using order parameters defined previously to obtain properties such as glazer tilt symbols, MA reorientation distributions, etc. These show excellent agreement with previous experiments and AIMD simulations. We also reproduce unique features of each of these phases, such as the triple-well behavior for octahedral tilts in the $\alpha$ phase.

Our simulation provide important new insights into these phases enabling us to fill gaps in the understanding of their behavior. For instance, in the $\beta$ phase, we observe the formation of two kinds of MAs indicated by previous experiments, unseen in previous AIMD simulations due to insufficient time scales and cell sizes. An order parameter capable of capturing this unique ordering was defined as well. The origin of this unique ordering was attributed to the octahedral tilt behavior of this phase which stabilizes two kinds of voids distinguished by the H-bonding interactions accommodated. These indicate that coupling of the host and guest in such systems can lead to unique short-range ordering behaviours.

Further, we also investigate the characteristics of polar domains in the high pressure $\gamma$ phase. Suitable order parameters which can distinguish between polar and anti-polar domains with different domains sizes are used to quantify the domain lifetimes. The decay times of domains span timescales greater than 50 ps, which would play important role in delaying carrier recombination lifetimes and improving photovoltaic properties. Interesting characteristics of domain lifetimes with pressure are revealed as well such as large lifetimes for domains of size 2, which decreases with pressure, indicating local stabilization of polar minima at lower pressures.

Our simulations emphasize the importance of cell sizes and simulation times to accurately capture the static and dynamic properties of such materials. These signify the increased relevance of machine learning force fields so as to determine materials properties and thereby guide the design of materials. We also emphasize that studies using MLFF to investigate pressure-induced transitions is rare and we are not aware of any such study on hybrid perovskites. Hence, this study paves a way for future studies on pressure-induced phenomena in similar systems.

\bibliography{reference}% Produces the bibliography via BibTeX.

@preamble{"\providecommand{\noopsort}[1]{}" #
   "\providecommand{\singleletter}[1]{#1}%"}

@article{dewolfTandems2023,
  title = {Tandems Have the Power},
  author = {De Wolf, Stefaan and Aydin, Erkan},
  year = {2023},
  month = jul,
  journal = {Science},
  volume = {381},
  number = {6653},
  pages = {30--31},
  issn = {0036-8075, 1095-9203},
  urldate = {2024-05-01},
  langid = {english}
}

@article{swainsonPhaseTransitionsPerovskite2003,
	title = {Phase Transitions in the Perovskite Methylammonium Lead Bromide, {{CH3ND3PbBr3}}},
	author = {Swainson, I.P. and Hammond, R.P. and Soulli{\`e}re, C. and Knop, O. and Massa, W.},
	year = 2003,
	month = nov,
	journal = {Journal of Solid State Chemistry},
	volume = {176},
	number = {1},
	pages = {97--104},
	issn = {00224596},
	doi = {10.1016/S0022-4596(03)00352-9},
	copyright = {https://www.elsevier.com/tdm/userlicense/1.0/},
	langid = {english}
}

@article{swainsonPressureResponseOrganic2007,
  title = {Pressure {{Response}} of an {{Organic}}---{{Inorganic Perovskite}}: {{Methylammonium Lead Bromide}}.},
  shorttitle = {Pressure {{Response}} of an {{Organic}}---{{Inorganic Perovskite}}},
  author = {Swainson, I. P. and Tucker, M. G. and Wilson, D. J. and Winkler, B. and Milman, V.},
  year = {2007},
  month = aug,
  journal = {ChemInform},
  volume = {38},
  number = {32},
  pages = {chin.200732007},
  issn = {0931-7597, 1522-2667},
  urldate = {2024-05-29},
  copyright = {http://onlinelibrary.wiley.com/termsAndConditions\#vor},
  langid = {english}
}

@article{wangPressureInducedPhaseTransformation2015a,
  title = {Pressure-{{Induced Phase Transformation}}, {{Reversible Amorphization}}, and {{Anomalous Visible Light Response}} in {{Organolead Bromide Perovskite}}},
  author = {Wang, Yonggang and L{\"u}, Xujie and Yang, Wenge and Wen, Ting and Yang, Liuxiang and Ren, Xiangting and Wang, Lin and Lin, Zheshuai and Zhao, Yusheng},
  year = {2015},
  month = sep,
  journal = {Journal of the American Chemical Society},
  volume = {137},
  number = {34},
  pages = {11144--11149},
  issn = {1520-5126},
  langid = {english},
  pmid = {26284441}
}

@article{jaffeHighPressureSingleCrystalStructures2016,
  title = {High-{{Pressure Single-Crystal Structures}} of {{3D Lead-Halide Hybrid Perovskites}} and {{Pressure Effects}} on Their {{Electronic}} and {{Optical Properties}}},
  author = {Jaffe, Adam and Lin, Yu and Beavers, Christine M. and Voss, Johannes and Mao, Wendy L. and Karunadasa, Hemamala I.},
  year = {2016},
  month = apr,
  journal = {ACS Central Science},
  volume = {2},
  number = {4},
  pages = {201--209},
  issn = {2374-7943, 2374-7951},
  urldate = {2024-01-27},
  langid = {english}
}

@article{capitaniLockingMethylammoniumPressureEnhanced2017,
  title = {Locking of {{Methylammonium}} by {{Pressure-Enhanced H-Bonding}} in ({{CH3NH3}}){{PbBr3 Hybrid Perovskite}}},
  author = {Capitani, F},
  year = {2017},
  journal = {The Journal of Physical Chemistry C}
}

@article{zhangEffectsNonhydrostaticStress2017,
  title = {Effects of {{Nonhydrostatic Stress}} on {{Structural}} and {{Optoelectronic Properties}} of {{Methylammonium Lead Bromide Perovskite}}},
  author = {Zhang, Rong and Cai, Weizhao and Bi, Tiange and Zarifi, Niloofar and Terpstra, Tyson and Zhang, Chuang and Verdeny, Z. Valy and Zurek, Eva and Deemyad, Shanti},
  year = {2017},
  month = aug,
  journal = {The Journal of Physical Chemistry Letters},
  volume = {8},
  number = {15},
  pages = {3457--3465},
  issn = {1948-7185},
  langid = {english},
  pmid = {28691486}
}

@article{yesudhasCouplingOrganicCation2020,
  title = {Coupling of Organic Cation and Inorganic Lattice in Methylammonium Lead Halide Perovskites: {{Insights}} into a Pressure-Induced Isostructural Phase Transition},
  shorttitle = {Coupling of Organic Cation and Inorganic Lattice in Methylammonium Lead Halide Perovskites},
  author = {Yesudhas, Sorb and Burns, Randy and Lavina, Barbara and Tkachev, Sergey N. and Sun, Jiuyu and Ullrich, Carsten A. and Guha, Suchismita},
  year = {2020},
  month = oct,
  journal = {Physical Review Materials},
  volume = {4},
  number = {10},
  pages = {105403},
  issn = {2475-9953},
  urldate = {2024-01-11},
  langid = {english}
}

@article{liangReassigningPressureInducedPhase2022,
  title = {Reassigning the {{Pressure-Induced Phase Transitions}} of {{Methylammonium Lead Bromide Perovskite}}},
  author = {Liang, Akun and {Gonzalez-Platas}, Javier and Turnbull, Robin and Popescu, Catalin and {Fernandez-Guillen}, Ismael and Abargues, Rafael and Boix, Pablo P. and Shi, Lan-Ting and Errandonea, Daniel},
  year = {2022},
  month = nov,
  journal = {Journal of the American Chemical Society},
  volume = {144},
  number = {43},
  pages = {20099--20108},
  issn = {0002-7863, 1520-5126},
  urldate = {2024-01-11},
  langid = {english}
}

@article{yinHighPressureInduced2018,
  title = {High-{{Pressure}}-{{Induced Comminution}} and {{Recrystallization}} of {{CH}} {\textsubscript{3}} {{NH}} {\textsubscript{3}} {{PbBr}} {\textsubscript{3}} {{Nanocrystals}} as {{Large Thin Nanoplates}}},
  author = {Yin, Tingting and Fang, Yanan and Chong, Wee Kiang and Ming, Koh Teck and Jiang, Shaojie and Li, Xianglin and Kuo, Jer-Lai and Fang, Jiye and Sum, Tze Chien and White, Timothy J. and Yan, Jiaxu and Shen, Ze Xiang},
  year = {2018},
  month = jan,
  journal = {Advanced Materials},
  volume = {30},
  number = {2},
  pages = {1705017},
  issn = {0935-9648, 1521-4095},
  urldate = {2024-05-31},
  copyright = {http://onlinelibrary.wiley.com/termsAndConditions\#vor},
  langid = {english}
}

@article{liangPressureInducedPhaseTransition2023,
  title = {Pressure-{{Induced Phase Transition}} versus {{Amorphization}} in {{Hybrid Methylammonium Lead Bromide Perovskite}}},
  author = {Liang, Akun and Turnbull, Robin and Popescu, Catalin and {Fernandez-Guillen}, Ismael and Abargues, Rafael and Boix, Pablo P. and Errandonea, Daniel},
  year = {2023},
  month = jul,
  journal = {The Journal of Physical Chemistry C},
  volume = {127},
  number = {26},
  pages = {12821--12826},
  issn = {1932-7447, 1932-7455},
  urldate = {2024-01-11},
  langid = {english}
}

@article{maityStabilizingPolarDomains2023,
  title = {Stabilizing {{Polar Domains}} in {{MAPbBr}} {\textsubscript{3}} via the {{Hydrostatic Pressure-Induced Liquid Crystal-like Transition}}},
  author = {Maity, Sayan and Verma, Suraj and Ramaniah, Lavanya M. and Srinivasan, Varadharajan},
  year = {2023},
  month = jun,
  journal = {The Journal of Physical Chemistry Letters},
  volume = {14},
  number = {24},
  pages = {5497--5504},
  issn = {1948-7185, 1948-7185},
  urldate = {2023-08-23},
  langid = {english}
}

@article{maityCooperativeOctahedralTilt2024a,
  title = {Cooperative {{Octahedral Tilt Modes Drive Coexisting Displacive}} and {{Order}}--{{Disorder Pressure-Induced Phase Transitions}} in {{MAPbBr}} {\textsubscript{3}}},
  author = {Maity, Sayan and Verma, Suraj and Ramaniah, Lavanya M. and Srinivasan, Varadharajan},
  year = {2024},
  month = may,
  journal = {The Journal of Physical Chemistry C},
  pages = {acs.jpcc.4c01857},
  issn = {1932-7447, 1932-7455},
  urldate = {2024-05-11},
  copyright = {https://doi.org/10.15223/policy-029},
  langid = {english}
}

@article{zengDeePMDkitV2Software2023,
  title = {{{DeePMD-kit}} v2: {{A}} Software Package for Deep Potential Models},
  shorttitle = {{{DeePMD-kit}} V2},
  author = {Zeng, Jinzhe and Zhang, Duo and Lu, Denghui and Mo, Pinghui and Li, Zeyu and Chen, Yixiao and Rynik, Mari{\'a}n and Huang, Li'ang and Li, Ziyao and Shi, Shaochen and Wang, Yingze and Ye, Haotian and Tuo, Ping and Yang, Jiabin and Ding, Ye and Li, Yifan and Tisi, Davide and Zeng, Qiyu and Bao, Han and Xia, Yu and Huang, Jiameng and Muraoka, Koki and Wang, Yibo and Chang, Junhan and Yuan, Fengbo and Bore, Sigbj{\o}rn L{\o}land and Cai, Chun and Lin, Yinnian and Wang, Bo and Xu, Jiayan and Zhu, Jia-Xin and Luo, Chenxing and Zhang, Yuzhi and Goodall, Rhys E. A. and Liang, Wenshuo and Singh, Anurag Kumar and Yao, Sikai and Zhang, Jingchao and Wentzcovitch, Renata and Han, Jiequn and Liu, Jie and Jia, Weile and York, Darrin M. and E, Weinan and Car, Roberto and Zhang, Linfeng and Wang, Han},
  year = {2023},
  month = aug,
  journal = {The Journal of Chemical Physics},
  volume = {159},
  number = {5},
  pages = {054801},
  issn = {0021-9606, 1089-7690},
  urldate = {2023-08-25},
  langid = {english}
}

@article{zhangDeepPotentialMolecular2018,
  title = {Deep {{Potential Molecular Dynamics}}: {{A Scalable Model}} with the {{Accuracy}} of {{Quantum Mechanics}}},
  shorttitle = {Deep {{Potential Molecular Dynamics}}},
  author = {Zhang, Linfeng and Han, Jiequn and Wang, Han and Car, Roberto and E, Weinan},
  year = {2018},
  month = apr,
  journal = {Physical Review Letters},
  volume = {120},
  number = {14},
  pages = {143001},
  issn = {0031-9007, 1079-7114},
  urldate = {2022-09-27},
  langid = {english}
}

@article{carUnifiedApproachMolecular1985,
  title = {Unified {{Approach}} for {{Molecular Dynamics}} and {{Density-Functional Theory}}},
  author = {Car, R. and Parrinello, M.},
  year = {1985},
  month = nov,
  journal = {Physical Review Letters},
  volume = {55},
  number = {22},
  pages = {2471--2474},
  issn = {0031-9007},
  urldate = {2024-05-31},
  copyright = {http://link.aps.org/licenses/aps-default-license},
  langid = {english}
}

@article{giannozziQUANTUMESPRESSOModular2009,
  title = {{{QUANTUM ESPRESSO}}: A Modular and Open-Source Software Project for Quantum Simulations of Materials},
  shorttitle = {{{QUANTUM ESPRESSO}}},
  author = {Giannozzi, Paolo and Baroni, Stefano and Bonini, Nicola and Calandra, Matteo and Car, Roberto and Cavazzoni, Carlo and Ceresoli, Davide and Chiarotti, Guido L and Cococcioni, Matteo and Dabo, Ismaila and Dal Corso, Andrea and De Gironcoli, Stefano and Fabris, Stefano and Fratesi, Guido and Gebauer, Ralph and Gerstmann, Uwe and Gougoussis, Christos and Kokalj, Anton and Lazzeri, Michele and {Martin-Samos}, Layla and Marzari, Nicola and Mauri, Francesco and Mazzarello, Riccardo and Paolini, Stefano and Pasquarello, Alfredo and Paulatto, Lorenzo and Sbraccia, Carlo and Scandolo, Sandro and Sclauzero, Gabriele and Seitsonen, Ari P and Smogunov, Alexander and Umari, Paolo and Wentzcovitch, Renata M},
  year = {2009},
  month = sep,
  journal = {Journal of Physics: Condensed Matter},
  volume = {21},
  number = {39},
  pages = {395502},
  issn = {0953-8984, 1361-648X},
  urldate = {2023-12-20}
}

@article{qiaoFerroelectricPolarizationSuppresses2019,
  title = {Ferroelectric {{Polarization Suppresses Nonradiative Electron}}--{{Hole Recombination}} in {{CH}} {\textsubscript{3}} {{NH}} {\textsubscript{3}} {{PbI}} {\textsubscript{3}} {{Perovskites}}: {{A Time-Domain}} Ab {{Initio Study}}},
  shorttitle = {Ferroelectric {{Polarization Suppresses Nonradiative Electron}}--{{Hole Recombination}} in {{CH}} {\textsubscript{3}} {{NH}} {\textsubscript{3}} {{PbI}} {\textsubscript{3}} {{Perovskites}}},
  author = {Qiao, Lu and Fang, Wei-Hai and Long, Run},
  year = {2019},
  month = nov,
  journal = {The Journal of Physical Chemistry Letters},
  volume = {10},
  number = {22},
  pages = {7237--7244},
  issn = {1948-7185, 1948-7185},
  urldate = {2024-01-13},
  langid = {english}
}

@article{zhang_deep_2018,
    title = {Deep {Potential} {Molecular} {Dynamics}: {A} {Scalable} {Model} with the {Accuracy} of {Quantum} {Mechanics}},
    volume = {120},
    issn = {0031-9007, 1079-7114},
    shorttitle = {Deep {Potential} {Molecular} {Dynamics}},
    url = {https://link.aps.org/doi/10.1103/PhysRevLett.120.143001},
    doi = {10.1103/PhysRevLett.120.143001},
    language = {en},
    number = {14},
    urldate = {2022-09-27},
    journal = {Physical Review Letters},
    author = {Zhang, Linfeng and Han, Jiequn and Wang, Han and Car, Roberto and E, Weinan},
    month = apr,
    year = {2018},
    pages = {143001},
}

@article{maity_deciphering_2022,
	title = {Deciphering the {Nature} of {Temperature}-{Induced} {Phases} of {MAPbBr} $_{\textrm{3}}$ by \textit{{Ab} {Initio}} {Molecular} {Dynamics}},
	volume = {34},
	issn = {0897-4756, 1520-5002},
	url = {https://pubs.acs.org/doi/abs/10.1021/acs.chemmater.2c02453},
	doi = {10.1021/acs.chemmater.2c02453},
	language = {en},
	number = {23},
	urldate = {2023-01-04},
	journal = {Chemistry of Materials},
	author = {Maity, Sayan and Verma, Suraj and Ramaniah, Lavanya M. and Srinivasan, Varadharajan},
	month = dec,
	year = {2022},
	pages = {10459--10469},
}

@article{dubajic_dynamic_2025,
	title = {Dynamic nanodomains dictate macroscopic properties in lead halide perovskites},
	issn = {1748-3387, 1748-3395},
	url = {https://www.nature.com/articles/s41565-025-01917-0},
	doi = {10.1038/s41565-025-01917-0},
	language = {en},
	urldate = {2025-06-12},
	journal = {Nature Nanotechnology},
	author = {Dubajic, Milos and Neilson, James R. and Klarbring, Johan and Liang, Xia and Bird, Stephanie A. and Rule, Kirrily C. and Auckett, Josie E. and Selby, Thomas A. and Tumen-Ulzii, Ganbaatar and Lu, Yang and Jung, Young-Kwang and Chosy, Cullen and Wei, Zimu and Boeije, Yorrick and Zimmermann, Martin V. and Pusch, Andreas and Gu, Leilei and Jia, Xuguang and Wu, Qiyuan and Trowbridge, Julia C. and Mozur, Eve M. and Minelli, Arianna and Roth, Nikolaj and Orr, Kieran W. P. and Mahboubi Soufiani, Arman and Kahmann, Simon and Kabakova, Irina and Ding, Jianning and Wu, Tom and Conibeer, Gavin J. and Bremner, Stephen P. and Nielsen, Michael P. and Walsh, Aron and Stranks, Samuel D.},
	month = jun,
	year = {2025},
}

@article{tuo_spontaneous_2023,
	title = {Spontaneous {Hybrid} {Nano}‐{Domain} {Behavior} of the {Organic}–{Inorganic} {Hybrid} {Perovskites}},
	issn = {1616-301X, 1616-3028},
	url = {https://onlinelibrary.wiley.com/doi/10.1002/adfm.202301663},
	doi = {10.1002/adfm.202301663},
	language = {en},
	urldate = {2023-06-21},
	journal = {Advanced Functional Materials},
	author = {Tuo, Ping and Li, Lei and Wang, Xiaoxu and Chen, Jianhui and Zhong, Zhicheng and Xu, Bo and Dai, Fu‐Zhi},
	month = apr,
	year = {2023},
	pages = {2301663},
}

@article{liang_structural_2023,
	title = {Structural {Dynamics} {Descriptors} for {Metal} {Halide} {Perovskites}},
	volume = {127},
	issn = {1932-7447, 1932-7455},
	url = {https://pubs.acs.org/doi/10.1021/acs.jpcc.3c03377},
	doi = {10.1021/acs.jpcc.3c03377},
	language = {en},
	number = {38},
	urldate = {2023-12-19},
	journal = {The Journal of Physical Chemistry C},
	author = {Liang, Xia and Klarbring, Johan and Baldwin, William J. and Li, Zhenzhu and Csányi, Gábor and Walsh, Aron},
	month = sep,
	year = {2023},
	pages = {19141--19151},
}

@article{maczkaPhaseTransitionsDielectric2023,
	title = {Phase {{Transitions}}, {{Dielectric Response}}, and {{Nonlinear Optical Properties}} of {{Aziridinium Lead Halide Perovskites}}},
	author = {M{\k a}czka, Miros{\l}aw and Ptak, Maciej and G{\k a}gor, Anna and Zar{\k e}ba, Jan K. and Liang, Xia and Bal{\v c}i{$\overline u$}nas, Sergejus and Semenikhin, Oleksandr A. and Kucheriv, Olesia I. and Gural'skiy, Il'ya A. and Shova, Sergiu and Walsh, Aron and Banys, J{$\overline u$}ras and {\v S}im{\.e}nas, Mantas},
	year = {2023},
	month = nov,
	journal = {Chemistry of Materials},
	volume = {35},
	number = {22},
	pages = {9725--9738},
	issn = {0897-4756, 1520-5002},
	doi = {10.1021/acs.chemmater.3c02200},
	langid = {english}
}

@article{fykouras_disorder_2023,
	title = {Disorder to order: how halide mixing in {MAPbI} $_{\textrm{ 3− \textit{x} }}$ {Br} $_{\textrm{ \textit{x} }}$ perovskites restricts {MA} dynamics},
	volume = {11},
	issn = {2050-7488, 2050-7496},
	shorttitle = {Disorder to order},
	url = {http://xlink.rsc.org/?DOI=D2TA09069D},
	doi = {10.1039/D2TA09069D},
	language = {en},
	number = {9},
	urldate = {2023-12-25},
	journal = {Journal of Materials Chemistry A},
	author = {Fykouras, Kostas and Lahnsteiner, Jonathan and Leupold, Nico and Tinnemans, Paul and Moos, Ralf and Panzer, Fabian and De Wijs, Gilles A. and Bokdam, Menno and Grüninger, Helen and Kentgens, Arno P. M.},
	year = {2023},
	pages = {4587--4597},
}

@article{fransson_revealing_2023,
	title = {Revealing the {Free} {Energy} {Landscape} of {Halide} {Perovskites}: {Metastability} and {Transition} {Characters} in {CsPbBr} $_{\textrm{3}}$ and {MAPbI} $_{\textrm{3}}$},
	volume = {35},
	issn = {0897-4756, 1520-5002},
	shorttitle = {Revealing the {Free} {Energy} {Landscape} of {Halide} {Perovskites}},
	url = {https://pubs.acs.org/doi/10.1021/acs.chemmater.3c01740},
	doi = {10.1021/acs.chemmater.3c01740},
	language = {en},
	number = {19},
	urldate = {2023-12-25},
	journal = {Chemistry of Materials},
	author = {Fransson, Erik and Rahm, J. Magnus and Wiktor, Julia and Erhart, Paul},
	month = oct,
	year = {2023},
	pages = {8229--8238},
}

@article{bokdam_exploring_2021,
	title = {Exploring {Librational} {Pathways} with on-the-{Fly} {Machine}-{Learning} {Force} {Fields}: {Methylammonium} {Molecules} in {MAPbX} $_{\textrm{3}}$ ({X} = {I}, {Br}, {Cl}) {Perovskites}},
	volume = {125},
	issn = {1932-7447, 1932-7455},
	shorttitle = {Exploring {Librational} {Pathways} with on-the-{Fly} {Machine}-{Learning} {Force} {Fields}},
	url = {https://pubs.acs.org/doi/10.1021/acs.jpcc.1c06835},
	doi = {10.1021/acs.jpcc.1c06835},
	language = {en},
	number = {38},
	urldate = {2023-09-07},
	journal = {The Journal of Physical Chemistry C},
	author = {Bokdam, Menno and Lahnsteiner, Jonathan and Sarma, D. D.},
	month = sep,
	year = {2021},
	pages = {21077--21086},
}

@article{jinnouchi_phase_2019,
	title = {Phase {Transitions} of {Hybrid} {Perovskites} {Simulated} by {Machine}-{Learning} {Force} {Fields} {Trained} on the {Fly} with {Bayesian} {Inference}},
	volume = {122},
	issn = {0031-9007, 1079-7114},
	url = {https://link.aps.org/doi/10.1103/PhysRevLett.122.225701},
	doi = {10.1103/PhysRevLett.122.225701},
	language = {en},
	number = {22},
	urldate = {2023-05-01},
	journal = {Physical Review Letters},
	author = {Jinnouchi, Ryosuke and Lahnsteiner, Jonathan and Karsai, Ferenc and Kresse, Georg and Bokdam, Menno},
	month = jun,
	year = {2019},
	pages = {225701},
}

@article{bernasconiDirectEvidencePermanent2017,
	title = {Direct {{Evidence}} of {{Permanent Octahedra Distortion}} in {{MAPbBr}}{\textsubscript{3}} {{Hybrid Perovskite}}},
	author = {Bernasconi, Andrea and Malavasi, Lorenzo},
	date = {2017-04-14},
	journaltitle = {ACS Energy Letters},
	shortjournal = {ACS Energy Lett.},
	volume = {2},
	number = {4},
	pages = {863--868},
	issn = {2380-8195, 2380-8195},
	doi = {10.1021/acsenergylett.7b00139},
	langid = {english}
}

@article{pageShortRangeOrderMethylammonium2016,
	title = {Short‐{{Range Order}} of {{Methylammonium}} and {{Persistence}} of {{Distortion}} at the {{Local Scale}} in {{MAPbBr}}{\textsubscript{3}} {{Hybrid Perovskite}}},
	author = {Page, Katharine and Siewenie, Joan E. and Quadrelli, Paolo and Malavasi, Lorenzo},
	date = {2016-11-07},
	journaltitle = {Angewandte Chemie International Edition},
	shortjournal = {Angew Chem Int Ed},
	volume = {55},
	number = {46},
	pages = {14320--14324},
	issn = {1433-7851, 1521-3773},
	doi = {10.1002/anie.201608602},
	langid = {english}
}

@article{glazerClassificationTiltedOctahedra1972,
	title = {The Classification of Tilted Octahedra in Perovskites},
	author = {Glazer, A. M.},
	year = 1972,
	month = nov,
	journal = {Acta Crystallographica Section B Structural Crystallography and Crystal Chemistry},
	volume = {28},
	number = {11},
	pages = {3384--3392},
	issn = {05677408},
	doi = {10.1107/S0567740872007976},
	copyright = {http://journals.iucr.org/services/copyrightpolicy.html}
}

@article{zhangDPGENConcurrentLearning2020,
	title = {{{DP-GEN}}: {{A}} Concurrent Learning Platform for the Generation of Reliable Deep Learning Based Potential Energy Models},
	shorttitle = {{{DP-GEN}}},
	author = {Zhang, Yuzhi and Wang, Haidi and Chen, Weijie and Zeng, Jinzhe and Zhang, Linfeng and Wang, Han and E, Weinan},
	year = 2020,
	month = aug,
	journal = {Computer Physics Communications},
	volume = {253},
	pages = {107206},
	issn = {00104655},
	doi = {10.1016/j.cpc.2020.107206},
	langid = {english}
}

@article{thompsonLAMMPSFlexibleSimulation2022,
	title = {{{LAMMPS}} - a Flexible Simulation Tool for Particle-Based Materials Modeling at the Atomic, Meso, and Continuum Scales},
	author = {Thompson, Aidan P. and Aktulga, H. Metin and Berger, Richard and Bolintineanu, Dan S. and Brown, W. Michael and Crozier, Paul S. and In 'T Veld, Pieter J. and Kohlmeyer, Axel and Moore, Stan G. and Nguyen, Trung Dac and Shan, Ray and Stevens, Mark J. and Tranchida, Julien and Trott, Christian and Plimpton, Steven J.},
	year = 2022,
	month = feb,
	journal = {Computer Physics Communications},
	volume = {271},
	pages = {108171},
	issn = {00104655},
	doi = {10.1016/j.cpc.2021.108171},
	langid = {english}
}

\end{document}